\documentclass[preprint]{elsarticle}

\usepackage{amsmath,amsfonts,amssymb}

\usepackage{microtype}

\usepackage[utf8]{inputenc}
\usepackage{graphicx}
\usepackage{color}

\newtheorem{thm}{Theorem}
\newtheorem{conj}{Conjecture}

\newcommand{\tr}{\mathop{\mathrm{tr}}}

\begin{document}

\title{The phase space structure of the oligopoly dynamical system by means of Darboux integrability
}

\author[ieuj,csrc]{Adam Krawiec}
\ead{adam.krawiec@uj.edu.pl}

\author[pan,amp]{Tomasz Stachowiak}
\ead{tomasz@amp.i.kyoto-u.ac.jp}

\author[oauj,csrc]{Marek Szyd{\l}owski}
\ead{marek.szydlowski@uj.edu.pl}

\address[ieuj]{Institute of Economics, Finance and Management, Jagiellonian University, Lojasiewicza 4, 30-348 Krak{\'o}w, Poland}

\address[pan]{Center for Theoretical Physics, Polish Academy of Sciences, Aleja Lotnik{\'o}w 32/46, Warszawa, Poland}
\address[amp]{Department of Applied Mathematics and Physics, Graduate School of
Informatics, Kyoto University, 606-8501 Kyoto, Japan}

\address[oauj]{Astronomical Observatory, Jagiellonian University, Orla 171, 30-244 Krak{\'o}w, Poland}

\address[csrc]{Mark Kac Complex Systems Research Centre, Jagiellonian University, Krak{\'o}w, Poland}

\date{}


\begin{abstract}
We investigate the dynamical complexity of Cournot oligopoly dynamics of three
firms by using the qualitative methods of dynamical systems to study the phase
structure of this model. The phase space is organized with one-dimensional and
two-dimensional invariant submanifolds (for the monopoly and duopoly) and
unique stable node (global attractor) in the positive quadrant of the phase
space (Cournot equilibrium). We also study the integrability of the system. We
demonstrate the effectiveness of the method of the Darboux polynomials in
searching for first integrals of the oligopoly. The general method as well as
examples of adopting this method are presented. We study Darboux
non-integrability of the oligopoly for linear demand functions and find first
integrals of this system for special classes of the system, in particular,
rational integrals can be found for a quite general set of model parameters.
We show how first integral can be useful in lowering the dimension of the
system using the example of $n$ almost identical firms.  This first integral
also gives information about the structure of the phase space and the behaviour
of trajectories in the neighbourhood of a Nash equilibrium.
\end{abstract}

\maketitle

\section{Introduction}

In the economic study of imperfect markets a special place is held by the
oligopoly. It is a market structure where a substantial market share is held by
a small number of firms. The behaviour of firms in the market, studied
first by Cournot \cite{Cournot:1838rpm}, consists in the firms producing a homogeneous
product and fixing the price taking into account that any change of price by a rival
firm influences its profit. There is an equilibrium in which these firms
maximize profits (it is a Nash equilibrium). The special cases of two and three
firms are called a duopoly and triopoly, respectively.

From the economic point of view the study of oligopoly dynamics is important as
we want to know a mechanism governing the market dynamics. The existence of
equilibrium of the market, its genericity, its stability and the dynamical
behaviour in the neighbourhood of the equilibrium are crucial for understanding
the formation and evolution of markets. In modern economics, the studies of
imperfect markets, especially the oligopoly model, is important as it is the
prevailing market structure. Accordingly, the oligopoly has been a
subject of intensive studies \cite{Puu:2011ooe}. Many researchers have developed different variants of the classical Cournot model. These investigations pursued both discrete and continuous time scales using the methods of dynamical systems. These studies concentrated on the asymptotic stability. The summary of earlier results can be found in Okuguchi \cite{Okuguchi:1976} while Bischi et al. presented comprehensive results for dynamics of discrete oligopoly systems \cite{Bischi:2010}.

In our analysis of the oligopoly, we consider two main problems.
First, we study the structure of the phase
space, by modelling the oligopoly as an autonomous $n$
dimensional system of ordinary differential equations.
The motivation for such investigation is looking for dynamical complexity of
the trajectories' behaviour in the phase space. Exploration of the phase
structure seems to be interesting because our knowledge about trajectories
describes the possible (long-term) behaviour of the firms and equilibria which
they reach.

The second is the problem of the existence of first integrals of a differential
equation systems. It has been studied in many dynamical systems of physical and
biological origin. However, in economic theory the problem of finding first
integrals of dynamical systems is almost absent. Among the few examples there
are the production/inventory model \cite{Steindl2004469} and the
multiplier-accelerator model of business cycle \cite{Valls20121573}. In this
paper we consider this problem for the oligopoly model.

To this aim, we propose to apply methods of the Darboux polynomials in finding
first and the so called ``second'' integrals. We then search for rational and
algebraic, time-dependent first integrals by means of combinations of the
Darboux polynomials.  Given enough such quantities, we can also obtain
time-independent first integrals of the original oligopoly system. The knowledge of
time-independent and functionally independent first integral yields important
information about phase space structure of any dynamical model.

If the system is two dimensional (duopoly)  then the existence of one first
integral means that the system is completely integrable and thus phase space
structure is completely characterized. If the first integral does  not depend
qualitatively on the special value of the model's parameters, then its phase
portrait is completely determined on the phase plane. In this case one can
easily study periodic orbits, limit cycles etc. 

The advantages of dynamical system methods is that they gives us the
possibility of geometrization of the dynamical behaviour and thus visualisation
of dynamics. The evolutional paths of a model are represented by trajectories
of a system. The phase space contains all trajectories for all admissible
initial conditions. Thus, the global information on dynamics is obtained. In
the phase space, trajectories can behave in a regular way or exhibit complex
behaviour \cite{Chian:2007csa}. The phase space has a structure organised by
fixed points, periodic orbits, invariant submanifolds, etc. The main aim of
this paper is reveal this structure in the context of oligopoly dynamics. To
this aim, we construct 3-dimensional phase portraits representative for the
problem and discuss in details integrability of the oligopoly systems. In the
discussion of problem of integrability we pay attention to Darboux
integrability. The analysis of integrability gives us information about
existence or not-existence of first integral. For visualization we construct
the level sets of the first integral. 

The problem of complex oligopoly dynamics was investigated in 
\cite{Puu:2005cod} for discrete version of oligopoly. In the paper we consider
smooth dynamical systems with continuous time. The existence of first integrals
gives us information on non-existence of complex chaotic behaviour. Although it
is not possible to formulate such conclusions in full generality, it holds for
sufficiently different firms. The generic nature of the critical points points
towards preservation of the regular phase space structure in the presence of
small perturbations. When they exist.

For 3-dimensional dynamical systems the information that there is first
integral means
that solutions are restricted to the level sets of such a function. If an
$n$-dimensional system admits a time independent first integral (as for example
in the case of oligopoly dynamics of $n$ almost identical firms), then the
first integral can be used to lower the dimension of the original system by one
(of course if we can effectively solve the conservation law for one of the
variables, so that it can be excluded from the system). We will illustrate how
for 3 almost identical firms the first integral can be used to lower the
dimension of the system to dimension two.

\section{The oligopoly model}

\subsection{General model}

Let us consider the oligopoly market with three firms. These three players
produce a homogeneous product of the total supply
$Q(t)=q_{1}(t)+q_{2}(t)+q_{3}(t)$. The price of the good is $p$ and the demand
is given by a linear inverse demand function
\begin{equation}\label{eq:1}
p(Q(t)) = a - bQ(t)
\end{equation}
where $a$ and $b$ are positive constants. The former is the highest market
price of the good. We assume that the firms' cost function has the quadratic
form
\begin{equation}\label{eq:2}
C_{i}(q_{i}) = c_{1} + d_{i}q_{i}(t) + e_{i}q_{i}^{2}(t), \quad i=1,2,3,
\end{equation}
where $c_i$ is the positive fixed cost of firm $i$, and $d_i$, $e_i$ are
constants. As the marginal cost of the firm must be less than the highest market
price we have the condition $d_{i} + 2e_{i}q_{i} < a$, $i=1,2,3$.

The profit of the $i$-th firm is
\begin{equation}\label{eq:3}
\Pi_{i}(q_{1}(t),q_{2}(t),q_{3}(t)) = q_{i}(t)(a-bQ(t)) 
- (c + d_{i}q_{i}(t) + e_{i}q_{i}^{2}(t)), \quad i=1,2,3,
\end{equation}
and its marginal profit is
\begin{equation}\label{eq:4}
\frac{\partial \Pi_{i}(q_{1},q_{2},q_{3})}{\partial q_{i}} 
= a - bQ(t) - bq_{i}(t) - d_{i} - 2e_{i}q_{i}(t), \quad i=1,2,3.
\end{equation}

We assume that all the firms have imperfect knowledge of the market. Therefore,
they follow a bounded rationality adjustment process based on a local estimate
of the marginal profit $\partial \Pi_{i}/\partial q_{i}$
\cite{Bischi:2000ga,Agiza:2002cd}. It means that the firm increases its
production as long as the marginal profit is positive. When the marginal profit
is negative the firm reduces its production. This adjustment mechanism has the
form
\begin{equation}\label{eq:5}
\frac{dq_i(t)}{dt}= \alpha_i q_{i}(t) 
\frac{\partial \Pi_{i}(q_{1}(t),q_{2}(t),q_{3}(t))}{\partial q_{i}}, 
\quad i=1,2,3,
\end{equation}
where $\alpha_{i}$ is a positive speed of adjustment. The three-firm oligopoly
is given by the system of differential equations
\begin{align}
\dot{q}_{1}(t) &= 
\alpha_1 q_{1}(t)[a - bQ(t) - bq_{1}(t) - d_{1} - 2e_{1}q_{1}(t)], 
\nonumber \\
\dot{q}_{2}(t) &= 
\alpha_2 q_{2}(t)[a - bQ(t) - bq_{2}(t) - d_{2} - 2e_{2}q_{2}(t)], 
\label{eq:6} \\
\dot{q}_{3}(t) &= 
\alpha_3 q_{3}(t)[a - bQ(t) - bq_{3}(t) - d_{3} - 2e_{3}q_{3}(t)],
\nonumber
\end{align}
or
\begin{align}
\dot{q}_{1}(t) &= 
\alpha_1 q_{1}(t)[a_{1} - 2 b_{1} q_{1}(t) - b q_{2}(t) - b q_{3}(t)], 
\nonumber \\
\dot{q}_{2}(t) &= 
\alpha_2 q_{2}(t)[a_{2} - b q_{1}(t) - 2 b_{2} q_{2}(t) - b q_{3}(t)], 
\label{eq:7} \\
\dot{q}_{3}(t) &= 
\alpha_3 q_{3}(t)[a_{3} - b q_{1}(t) - b q_{2}(t) - 2 b_{3} q_{3}(t)],
\nonumber
\end{align}
where
\begin{align}
a_{1} &= a - d_{1}, \qquad a_{2} = a - d_{2}, \qquad a_{3} = a - d_{3}, \nonumber \\
b_{1} &= b + e_{1}, \qquad b_{2} = b + e_{2}, \qquad b_{3} = b + e_{3}. \label{eq:8}
\end{align}

The variables $q_{i}$, $i=1,2,3$ must be non-negative to be economically
significant. Below we list more explicitly some simple sub-cases of interest.

\subsection{Identical firms with linear cost function}

Let the cost function be linear and identical for all three firms
\begin{equation}
C(q_i) = c + d q_i, \quad c>0, d>0
\end{equation}
Additionally, we assume that the speed of adjustment $\alpha$ is the same for
all three firms so without loss of generality we take $\alpha = 1$, then the
dynamical system has the form
\begin{align}
\dot{q}_{1}(t) &= 
q_{1}(t)[\bar{a} - 2 b q_{1}(t) - b q_{2}(t) - b q_{3}(t)], 
\nonumber \\
\dot{q}_{2}(t) &= 
q_{2}(t)[\bar{a} - b q_{1}(t) - 2 b q_{2}(t) - b q_{3}(t)], 
\label{eq:7_lc} \\
\dot{q}_{3}(t) &= 
q_{3}(t)[\bar{a} - b q_{1}(t) - b q_{2}(t) - 2 b q_{3}(t)],
\nonumber
\end{align}
where
\[
\bar{a} = a - d > 0, \quad a>0, \quad d>0 \quad \text{and} \quad b > 0.
\]

\subsection{Identical firms with quadratic cost function}

Now the cost function is quadratic
\begin{equation}
C(q_i) = c + d q_i + e q_{i}^{2}, \quad c>0, d>0, e>0.
\end{equation}
Additionally, we assume that the speed of adjustment $\alpha$ is the same for
all three firms so without loss of generality we take $\alpha = 1$, then the
dynamical system has the form
\begin{align}
\dot{q}_{1}(t) &= 
q_{1}(t)[\bar{a} - 2 \bar{b} q_{1}(t) - b q_{2}(t) - b q_{3}(t)], 
\nonumber \\
\dot{q}_{2}(t) &= 
q_{2}(t)[\bar{a} - b q_{1}(t) - 2 \bar{b} q_{2}(t) - b q_{3}(t)], 
\label{eq:7b} \\
\dot{q}_{3}(t) &= 
q_{3}(t)[\bar{a} - b q_{1}(t) - b q_{2}(t) - 2 \bar{b} q_{3}(t)],
\nonumber
\end{align}
where
\[
\bar{a} = a - d > 0 \quad \text{and} \quad \bar{b} = b + e > 0.
\]

\subsection{Different firms with quadratic cost function}

Now the cost function is quadratic and different for each firm
\begin{equation}
C_i(q_i) = c_i + d q_i + e_i q_{i}^{2}, \quad c_i>0, d_i>0, e_i>0
\end{equation}
Additionally, we assume that the speed of adjustment $\alpha$ is the same for
all three firms and without loss of generality we take $\alpha = 1$, then the
dynamical system has the form
\begin{align}
\dot{q}_{1}(t) &= 
q_{1}(t)[a_1 - 2 b_1 q_{1}(t) - b_2 q_{2}(t) - b_3 q_{3}(t)], 
\nonumber \\
\dot{q}_{2}(t) &= 
q_{2}(t)[a_2 - b q_{1}(t) - 2 b_2 q_{2}(t) - b q_{3}(t)], 
\label{eq:7c} \\
\dot{q}_{3}(t) &= 
q_{3}(t)[a_3 - b q_{1}(t) - b q_{2}(t) - 2 b_3 q_{3}(t)],
\nonumber
\end{align}
where
\begin{gather*}
a_i = a - d_i > 0 \quad \text{and} \quad b_i = b + e_i > 0, \\
a > 0, \quad d_i > 0, \quad b > 0, \quad e_i > 0. \label{eq:41b}
\end{gather*}

\section{Analysis of the model's dynamics}

\subsection{General model}
In this section we present the general analysis of dynamical system (\ref{eq:7})
\begin{align}
\dot{q}_{1}(t) &= 
\alpha_1 q_{1}(t)[a_{1} - 2 b_{1} q_{1}(t) - b q_{2}(t) - b q_{3}(t)] = h^1(q_1,q_2,q_3), 
\nonumber \\
\dot{q}_{2}(t) &= 
\alpha_2 q_{2}(t)[a_{2} - b q_{1}(t) - 2 b_{2} q_{2}(t) - b q_{3}(t)] = h^2(q_1,q_2,q_3), 
\label{eq:9} \\
\dot{q}_{3}(t) &= 
\alpha_3 q_{3}(t)[a_{3} - b q_{1}(t) - b q_{2}(t) - 2 b_{3} q_{3}(t)] = h^3(q_1,q_2,q_3),
\nonumber
\end{align}
where
\begin{align}
a_{1} &= a - d_{1}, \qquad a_{2} = a - d_{2}, \qquad a_{3} = a - d_{3}, \nonumber \\
b_{1} &= b + e_{1}, \qquad b_{2} = b + e_{2}, \qquad b_{3} = b + e_{3}. \label{eq:10}
\end{align}

To find a critical points of system (\ref{eq:7}) we solve the system
$h^{i}(q^*) = 0$, $i=1,2,3$, for $(q_1^*, q_2^*, q_3^*)$. Here, we
obtain multiple solutions:
\begin{align} \label{eq:20}
E_{1} &= (0,\ 0,\ 0) \\
E_{2} &= \left( \frac{a_1}{2b_1},\ 0,\ 0 \right) \\
E_{3} &= \left( 0,\ \frac{a_2}{2b_2},\ 0 \right) \\
E_{4} &= \left( 0,\ 0,\ \frac{a_3}{2b_3} \right) \\
E_{5} &= \left( \frac{2 a_1 b_2 - b a_2}{4 b_1 b_2 - b^2},\ 
\frac{2 a_2 b_1 - b a_1}{4 b_1 b_2 -b^2},\ 0 \right) \\
E_{6} &= \left( 0,\ \frac{2 a_2  b_3 - b a_3}{4 b_2 b_3 - b^2},\ 
\frac{2 a_3  b_2 - b a_2}{4 b_2  b_3 - b^2} \right) \\
E_{7} &= \left( \frac{2 a_1  b_3 - b a_3}{4 b_1 b_3 - b^2},\ 0,\
\frac{2 a_3  b_1 - b a_1}{4 b_1 b_3 - b^2} \right) \\
E_{8} &= \left( \frac{4a_1 b_2 b_3 + (-a_1 + a_2 + a_3)b^2 - 2(a_2 b_3 + a_3 b_2)b}{8b_1 b_2 b_3 + 2b^3 - 2(b_1 + b_2 + b_3)b^2}, \right. \nonumber \\
&\qquad \frac{4a_2 b_1 b_3 + (a_1 - a_2 + a_3)b^2 - 2(a_1 b_3 + a_3 b_1)b}{8b_1 b_2 b_3 + 2b^3 - 2(b_1 + b_2 + b_3)b^2}, \nonumber \\
&\qquad \left. \frac{4a_3 b_1 b_2 + (a_1 + a_2 - a_3)b^2 - 2(a_1 b_2 + a_2 b_1)b}{8b_1 b_2 b_3 + 2b^3 - 2(b_1 + b_2 + b_3)b^2} \right)
\end{align}

All these are points in the phase space which represent stationary
states of the system. The critical point $E_{1}$ is trivial without any supply
in the market. The critical points $E_{2}$, $E_{3}$ and $E_{4}$ correspond to a
monopoly market with firm ``1'', firm ``2'' or firm ``3'', respectively. The
critical points $E_{5}$, $E_{6}$ and $E_{7}$ correspond to a duopoly market
with firms ``1'' and ``2'', firms ``1'' and ``3'', or firms ``2'' and ``3'',
respectively. The last critical point $E_{8}$ corresponds to a three firms
oligopoly.

In the critical point $E_{8}$ all three firms choose their supply to be optimal
in such a way that their profit is maximized. It corresponds to the Cournot
equilibrium for three firms, and is also a Nash equilibrium as no firm 
wants to change (increase or decrease) its supply as this would lead to a
decrease in its profit.

There also exists a generic correspondence between the critical points and sets
invariant with respect to the flow.
The point $E_8$ usually lies outside invariant submanifolds, the monopolies
$E_2$, $E_3$, $E_4$ lie on one-dimensional submanifolds, the duopolies $E_5$,
$E_6$, $E_7$ lie on two dimensional submanifolds, and $E_1$ could be said to
lie on a zero-diemnsional one, but this is, strictly speaking, true for any
critical point.

From (\ref{eq:20}) we have that in the Cournot equilibrium the total supply is
\begin{equation}
Q^* = q_1^* + q_2^* + q_3^*
\end{equation}
and the price is
\begin{equation}
P^* = a - b(q_1^* + q_2^* + q_3^*).
\end{equation}

Let us try local stability analysis of the critical points represented the
Cournot equilibrium ($E_8$). Following the Hartman-Grobman theorem
\cite{Perko:2001de} the dynamics in the neighbourhood of this non-degenerate
critical point is approximated through linear part of the system.
The linearization at $q^*$ is
\begin{equation} \label{eq:21}
    \frac{d}{dt} \left[ \begin{array}{c}
    q_1 - q_1^* \\ q_2 - q_2^* \\ q_3 - q_3^*
    \end{array} \right] =
    M \left[ \begin{array}{c}
    q_1 - q_1^* \\ q_2 - q_2^* \\ q_3 - q_3^*
    \end{array} \right]
\end{equation}
where $M$ is the Jacobian calculated at $q^*$
\begin{align}
M &= \left[ \begin{array}{ccc}
    \frac{\partial h^1}{\partial q_1} \big|_{q=q^*} &
    \frac{\partial h^2}{\partial q_1} \big|_{q=q^*} &
    \frac{\partial h^3}{\partial q_1} \big|_{q=q^*} \\
    \frac{\partial h^1}{\partial q_2} \big|_{q=q^*} &
    \frac{\partial h^2}{\partial q_2} \big|_{q=q^*} &
    \frac{\partial h^3}{\partial q_2} \big|_{q=q^*} \\
    \frac{\partial h^1}{\partial q_3} \big|_{q=q^*} &
    \frac{\partial h^2}{\partial q_3} \big|_{q=q^*} &
    \frac{\partial h^3}{\partial q_3} \big|_{q=q^*}
    \end{array} \right] \\
    &= \left[ \begin{array}{ccc}
    -2 \alpha_1 b_1 q_1^* + \alpha_1 g^1 &
    -\alpha_1 b q_1^* & 
    -\alpha_1 b q_1^* \\
    -\alpha_2 b q_2^* &
    -2 \alpha_2 b_2 q_2^* + \alpha_2 g^2 &
    -\alpha_2 b q_2^* \\
    -\alpha_3 b q_3^* &
    -\alpha_3 b q_3^* &
    -2 \alpha_3 b_3 q_3^* + \alpha_3 g^3
    \end{array} \right]
\end{align}
where
\begin{align*}
g^1 = g^1(q_{1}^{*}, q_{2}^{*}, q_{3}^{*}) &= a_1 - 2 b_1 q_1^{*} - b q_2^{*} - b q_3^* \\
g^2 = g^2(q_{1}^{*}, q_{2}^{*}, q_{3}^{*}) &= a_1 - b q_1^* - 2 b_2 q_2^* - b q_3^* \\
g^3 = g^3(q_{1}^{*}, q_{2}^{*}, q_{3}^{*}) &= a_1 - b q_1^* - b q_2^* - 2 b_3 q_3^*.
\end{align*}

\subsection{Local stability analysis of the general model}

The stability of the critical point as well as its character depends on the eigenvalues of the linearization matrix which are the solutions of the characteristic equation
\begin{align}
\det [M - \lambda \mathbb{I}] &= \lambda^3 - \tr{M} \lambda^2 + ((\tr M)^2 - \tr M^2) \lambda \nonumber \\
\qquad &\quad + (\tr M)^3 + 2\tr M^3 - 3\tr M \tr M^2 \nonumber \\
\qquad &= m_3 \lambda^3 + m_2 \lambda_2 + m_1 \lambda + m_0= 0.
\end{align}

Let us consider critical point $E_8$. From the economic point of view this point
is the equilibrium for oligopoly of three firms. The Jacobi matrix calculated
at this critical point is
\begin{equation}
M = \left[ \begin{array}{ccc}
    -2 \alpha_1 b_1 q_1^* &
    -\alpha_1 b q_1^* & 
    -\alpha_1 b q_1^* \\
    -\alpha_2 b q_2^* &
    -2 \alpha_2 b_2 q_2^* &
    -\alpha_2 b q_2^* \\
    -\alpha_3 b q_3^* &
    -\alpha_3 b q_3^* &
    -2 \alpha_3 b_3 q_3^* 
    \end{array} \right]
\end{equation}
and the characteristic equation is
\begin{equation}
\begin{split}
\det[M-\lambda \mathbb{I}] = m_3 \lambda^3 + m_2 \lambda_2 + m_1 \lambda + m_0 = \\
\lambda^3 + 2(\alpha_1 b_1 q_1^* + \alpha_2 b_2 q_2^* + \alpha_3 b_3 q_3^*) \lambda^2 \\
+ [\alpha_1 \alpha_2 q_1^* q_2^* (4b_1 b_2 - b^2) + \alpha_1 \alpha_3 q_1^* q_3^* (4b_1 b_3 - b^2) + \alpha_2 \alpha_3 q_2^* q_3^* (4b_2 b_3 - b^2)] \lambda \\
+ 2 \alpha_1 \alpha_2 \alpha_2 q_1^* q_2^* q_3^* (b^3 + 4 b_1 b_2 b_3 - b^2 b_1 - b^2 b_2 - b^2 b_3) = 0
\end{split}
\end{equation}

The sign of the discriminant $\Delta$ of the characteristic equation determines
whether eigenvalues are real or complex. If the discriminant is negative,
then all eigenvalues are real. In turn, from the Routh-Hurvitz stability
criterion we have that if all coefficients of characteristic equation are
positive $a_i > 0$ and the condition $a_2 a_1 > a_3 a_0$ is fulfilled, then
the critical point is stable. The first condition of the criterion is always fulfilled
because $b_i > b$, $i=1,2,3$ and all parameters $c_i$, $i=0,1,2,3$ are
positive.

To deal with the second condition, we consider three possible positions of the
point $E_8$ which give additional inequalities for the parameters
constraints
\begin{itemize}
\item there is a critical point inside the positive quadrant of the phase space
(point $E_8$)
\[
\text{case I} \qquad q_{i}^* > 0 \quad i=1,2,3.
\]
\item there is no critical point inside the positive quadrant of the phase
    space; the critical point $E_8$ is located on a 2-dimensional invariant
        manifold
\[
\text{case IIa} \qquad q_i^* > 0, \qquad q_j^* > 0, \qquad q_k^* = 0,
        \quad i\neq j\neq k\neq i
\]
\item there is no critical point inside the positive quadrant of the phase; the
critical point $E_8$ is located on a 1-dimensional invariant manifold
\[
\text{case IIb} \qquad q_i^* > 0, \qquad q_j^* = q_k^* = 0, 
        \quad i\neq j\neq k\neq i. 
\]
\end{itemize}

\begin{figure}
    \centering
    \includegraphics[width=0.48\textwidth]{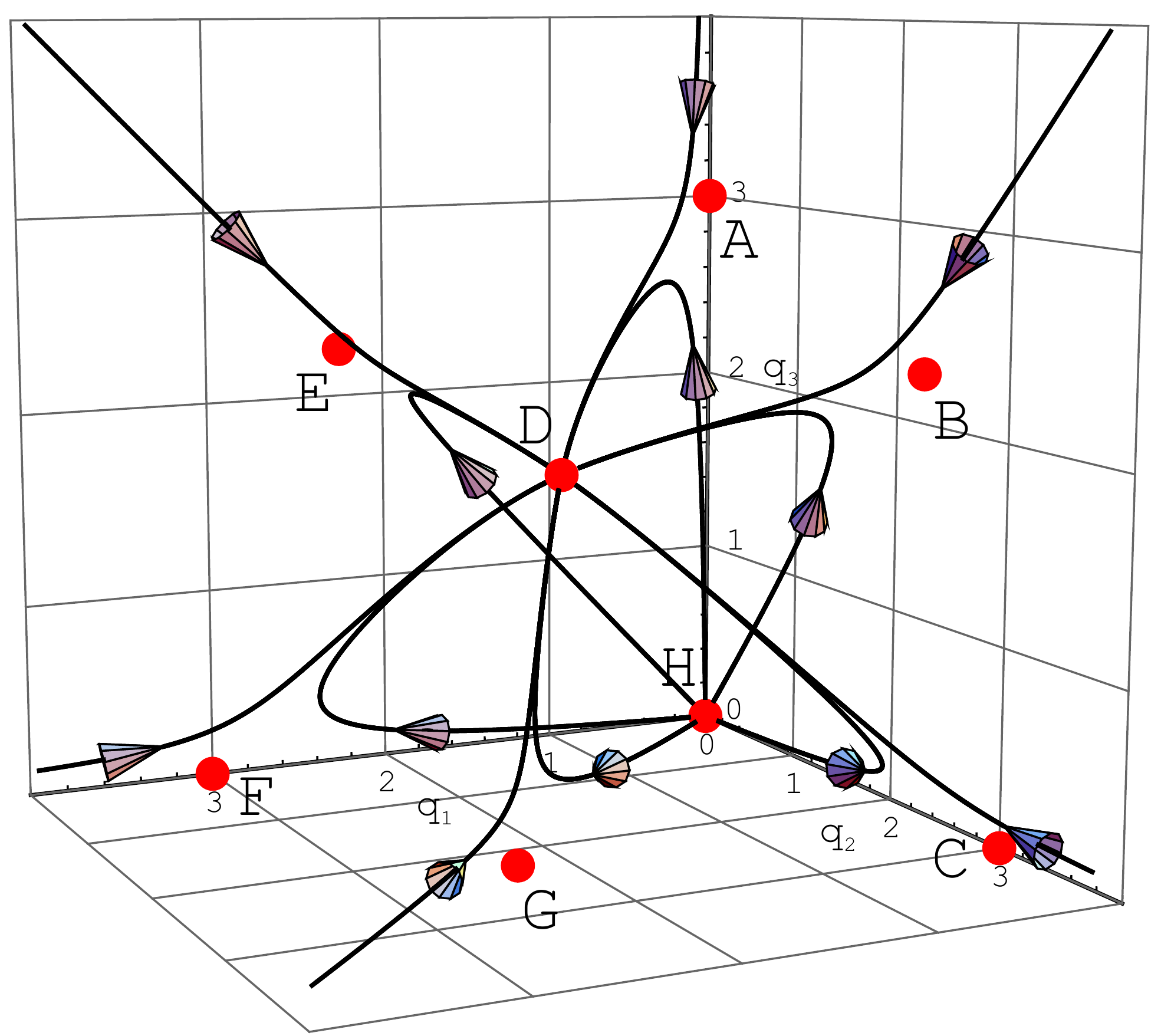}
    \includegraphics[width=0.48\textwidth]{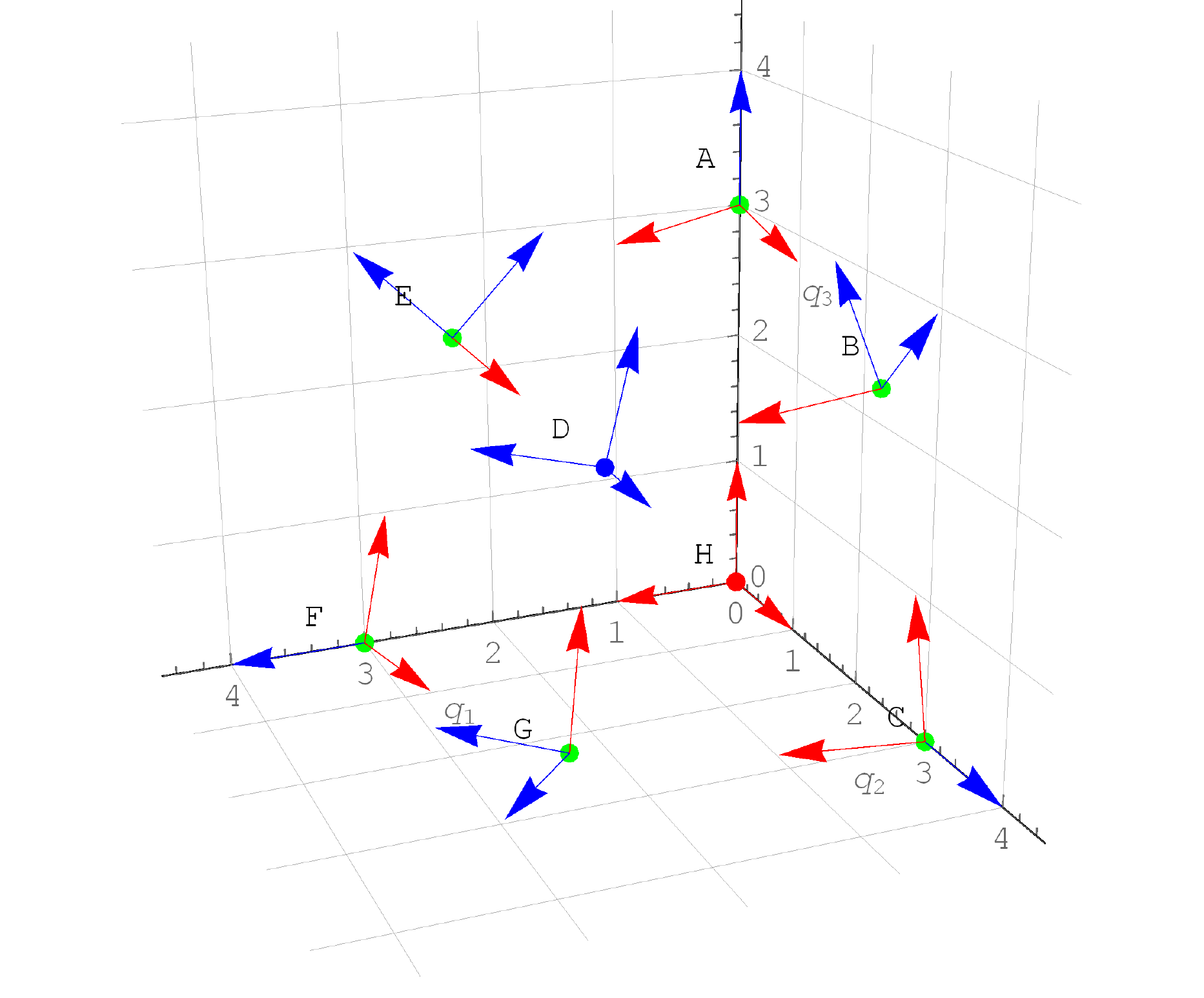}
    \caption{Left panel: the phase portrait for the system (\ref{eq:7}) with
the condition $4a_1 b_2 b_3 + (-a_1 + a_2 + a_3)b^2 - 2(a_2 b_3 + a_3 b_2)b >
0$, $4a_2 b_1 b_3 + (a_1 - a_2 + a_3)b^2 - 2(a_1 b_3 + a_3 b_1)b > 0$ and $4a_3
b_1 b_2 + (a_1 + a_2 - a_3)b^2 - 2(a_1 b_2 + a_2 b_1)b > 0$. Right panel: the
eigenvectors for these critical points.}
    \label{fig:1}
\end{figure}

For case I, the phase portrait of system (\ref{eq:7}) is presented in
fig.~\ref{fig:1} with conditions
\begin{equation}
\begin{aligned}
q_1^* &> 0 \quad \text{i.e.}\quad 4a_1 b_2 b_3 + (-a_1 + a_2 + a_3)b^2 
    - 2(a_2 b_3 + a_3 b_2)b > 0, \\
q_2^* &> 0 \quad \text{i.e.}\quad 4a_2 b_1 b_3 + (a_1 - a_2 + a_3)b^2 
    - 2(a_1 b_3 + a_3 b_1)b > 0, \\
q_3^* &> 0 \quad \text{i.e.}\quad 4a_3 b_1 b_2 + (a_1 + a_2 - a_3)b^2 
    - 2(a_1 b_2 + a_2 b_1)b > 0.
\end{aligned}
\end{equation}
The stable node $E_8$ is inside the positive quadrant of the phase space. Under
the above conditions, the set of initial conditions which lead to this point is
the whole positive quadrant. This point represents the Cournot equilibrium
where three firms, maximazing profits, coexist in the market. This is the
unique point in the positive quadrant which is a global attractor.

\begin{figure}
    \centering
    \includegraphics[width=0.48\textwidth]{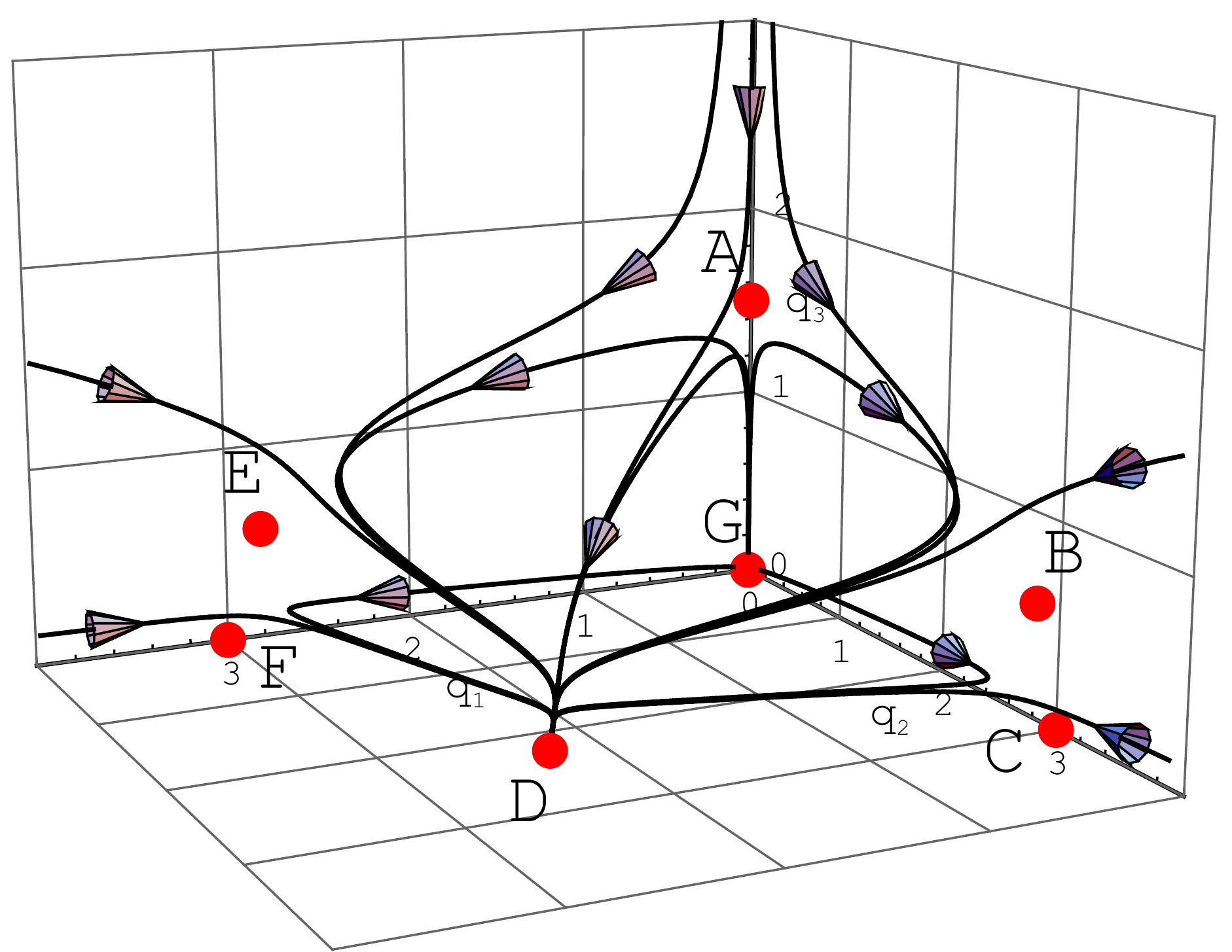}
    \includegraphics[width=0.48\textwidth]{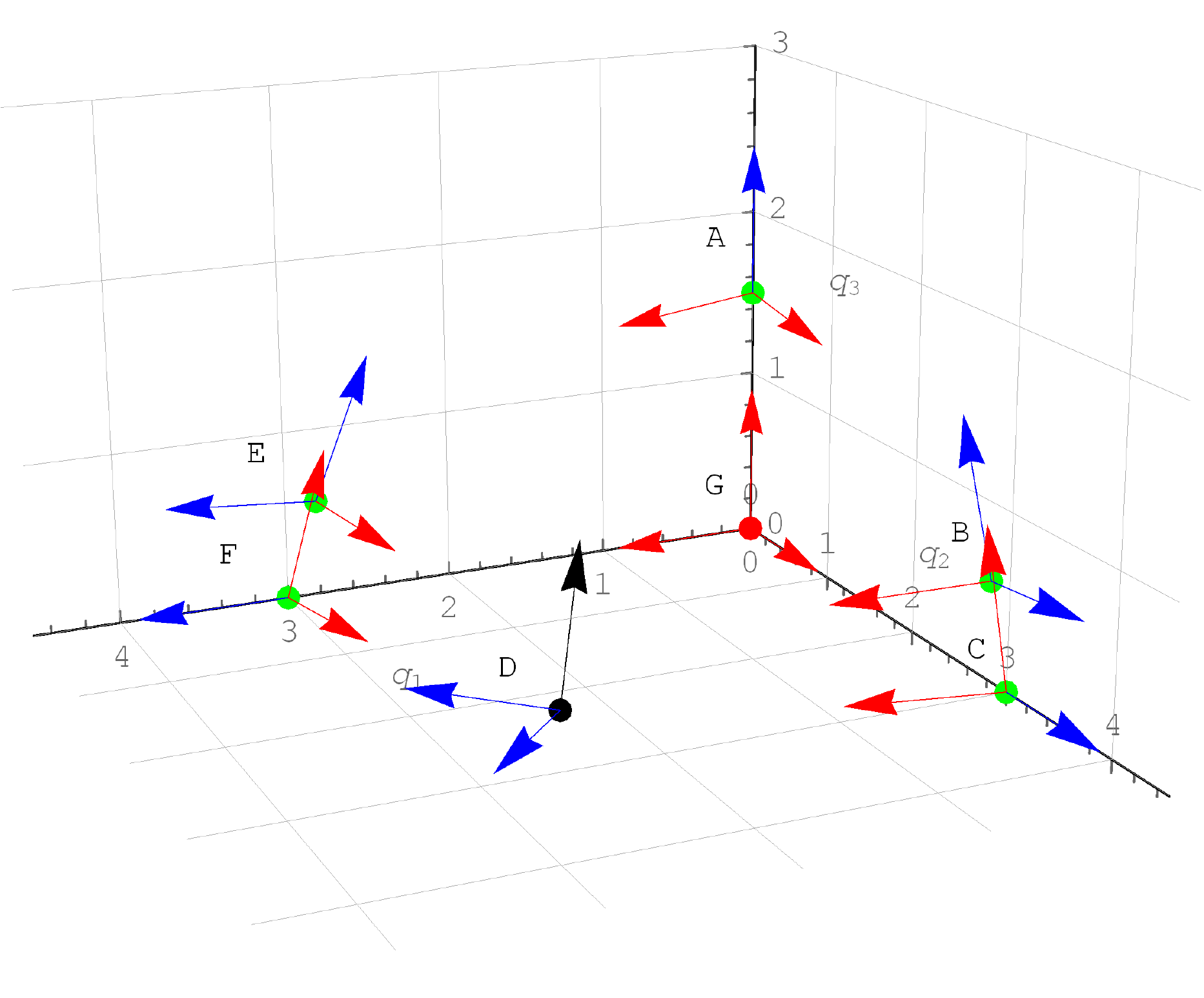}
    \caption{Left panel: The phase portrait for the system (\ref{eq:7}) with the conditions $4a_1 b_2 b_3 + (-a_1 + a_2 + a_3)b^2 - 2(a_2 b_3 + a_3 b_2)b > 0$, $4a_2 b_1 b_3 + (a_1 - a_2 + a_3)b^2 - 2(a_1 b_3 + a_3 b_1)b$ and $4a_3 b_1 b_2 + (a_1 + a_2 - a_3)b^2 - 2(a_1 b_2 + a_2 b_1)b = 0$. Right panel: the eigenvectors for these critical points.}
    \label{fig:2}
\end{figure}

For case IIa, the phase portrait of system (\ref{eq:7}) with conditions
\begin{equation}
\begin{aligned}
q_1^* &> 0 \quad \text{i.e.}\quad 4a_1 b_2 b_3 + (-a_1 + a_2 + a_3)b^2 - 2(a_2 b_3 + a_3 b_2)b > 0, \\
q_2^* &> 0 \quad \text{i.e.}\quad 4a_2 b_1 b_3 + (a_1 - a_2 + a_3)b^2 - 2(a_1 b_3 + a_3 b_1)b > 0, \\
q_3^* &= 0 \quad \text{i.e.}\quad 4a_3 b_1 b_2 + (a_1 + a_2 - a_3)b^2 - 2(a_1 b_2 + a_2 b_1)b = 0.
\end{aligned}
\end{equation}
is presented in fig.~\ref{fig:2}. In this case there is no critical point in
the positive quadrant of the phase space. The set of initial conditions
constituing the positive quadrant leads to the critical point located on the
two-dimensional invariant submanifold $q_3=0$. It means that the marginal cost of one of the firms
(paramater $d$) is sufficiently greater than marginal costs of other firms,
such that as time goes to infinity the firm reduces its production to zero.

This case describes evolutional scenario that one of the firms gradually
withdraws from the market due to costs higher than for the competition.

Case IIb is a variant of case IIa, where two firms have higher marginal costs
than the third firm. Both these firms gradually withdraw from market and only one
firm, the monopolist, is left.

For deeper understanding of the phase portrait, the additional map of
eigenvalues calculated at the critical points is presented in both previous
figures. These vectors span the plane tangent to the stable and unstable
manifolds.

To show the explicite implication of Routh-Hurvitz criterion, let us consider a simplier case of identical firms with a linear cost function and speed adjustment to market is equal 1, i.e.
\begin{equation}
    \alpha_1 = \alpha_2 = \alpha_3 = 1, \quad b_1 = b_2 = b_3 = b
\end{equation}
and the critical point $E_8$ has the coordinates
\begin{equation} \label{eq:s2_cp}
    q_1^* = q_2^* = q_3^* = q^* = \frac{\bar{a}}{4}
\end{equation}
where
$\bar{a} = a - d$ and $d = d_1 = d_2 = d_3$.
Now the characteristic equation has the form
\begin{equation} \label{eq:24_ce}
    \lambda^3 + \frac{3}{2} \bar{a} \lambda^2 + \frac{9}{16} \bar{a}^2 \lambda
    + \frac{1}{16} \bar{a}^3 = c_3 \lambda^3 + c_2 \lambda^2 + c_1 \lambda 
    + c_0 = 0.
\end{equation}

Let us state the theorem describing two possibilities for eigenvalues in the system (\ref{eq:7_lc}).
\begin{thm}
For system (\ref{eq:7_lc}) equilibrium point $E_8$ is a stable node.
\end{thm}

The discriminant of the third order polynomial (\ref{eq:24_ce}) is
\begin{equation} \label{eq:th}
\Delta = \frac{q^2}{4} + \frac{p^3}{27} = 0
\end{equation}
where
\[
p = - \frac{3}{16} \bar{a}^2, \qquad q = \frac{1}{32}\bar{a}^3.
\]
Both $p$ and $q$ are different from zero so then there is \\
a) one real eigenvalue of multiplicity one \\
b) and one real eigenvalue of multiplicity two. 

Let us check the Routh-Hurvitz criterion of stability of the critical point.
All the coefficients of the characteristic equation (\ref{eq:24_ce}) are positive. The condition 
$(m_2 m_1 > m_3 m_0)$ is also fulfilled for any parameter $\bar{a}$.

Hence, the critical point (\ref{eq:s2_cp}) is a stable node.

\subsection{Invariant submanifolds}

So far, the dynamics of the system was considered inside the phase space $\{
(q_1, q_2,q_3) \colon q_i > 0, i=1,2,3 \}$. Now, let us consider the
two-dimensional planes in this 3-dimensional phase space defined as $\{
(q_1,q_2), q_3=0\}$, $\{ (q_1,q_3), q_2=0\}$, and $\{ (q_2,q_3), q_1=0\}$.  The
system (\ref{eq:21}) possesses at least three invariant submanifolds on which
it assumes the form of two-dimensional autonomous dynamical system.

For the first invariant submanifold the system has the form
\begin{align}
\dot{q}_{1}(t) &= 0, 
\nonumber \\
\dot{q}_{2}(t) &= 
\alpha_2 q_{2}(t)[a_{2} - b q_{1}(t) - 2 b_{2} q_{2}(t) - b q_{3}(t)], 
\label{eq:7_s1} \\
\dot{q}_{3}(t) &= 
\alpha_3 q_{3}(t)[a_{3} - b q_{1}(t) - b q_{2}(t) - 2 b_{3} q_{3}(t)],
\nonumber
\end{align}

For the second invariant submanifold the system has the form
\begin{align}
\dot{q}_{1}(t) &= 
q_{1}(t)[a_{1} - 2 b_{1} q_{1}(t) - b q_{2}(t) - b q_{3}(t)], 
\nonumber \\
\dot{q}_{2}(t) &= 0, 
\label{eq:7_s2} \\
\dot{q}_{3}(t) &= 
\alpha_3 q_{3}(t)[a_{3} - b q_{1}(t) - b q_{2}(t) - 2 b_{3} q_{3}(t)],
\nonumber
\end{align}

For the third invariant submanifold the system has the form
\begin{align}
\dot{q}_{1}(t) &= 
q_{1}(t)[a_{1} - 2 b_{1} q_{1}(t) - b q_{2}(t) - b q_{3}(t)], 
\nonumber \\
\dot{q}_{2}(t) &= 
\alpha_2 q_{2}(t)[a_{2} - b q_{1}(t) - 2 b_{2} q_{2}(t) - b q_{3}(t)], 
\label{eq:7_s3} \\
\dot{q}_{3}(t) &= 0,
\nonumber
\end{align}
where $a_i = a - d_i$ and $b_i = b + e_i$.

These exist regardless of the values of the parameters, but we will be able to
find other linear submanifolds under some general assumptions -- we defer this
discussion to the section on Darboux polynomials.

Let us choose the following values of the parameters
\begin{equation}
\alpha_1 = 1,\; a_1 = 10,\; a_2 = 20,\; a_3 = 30,\;
b=0.5,\; e_i = 0.
\label{eq:33}
\end{equation}
to present the dynamics on the invariant submanifolds.

\begin{figure}
    \centering
    \includegraphics[scale=0.7]{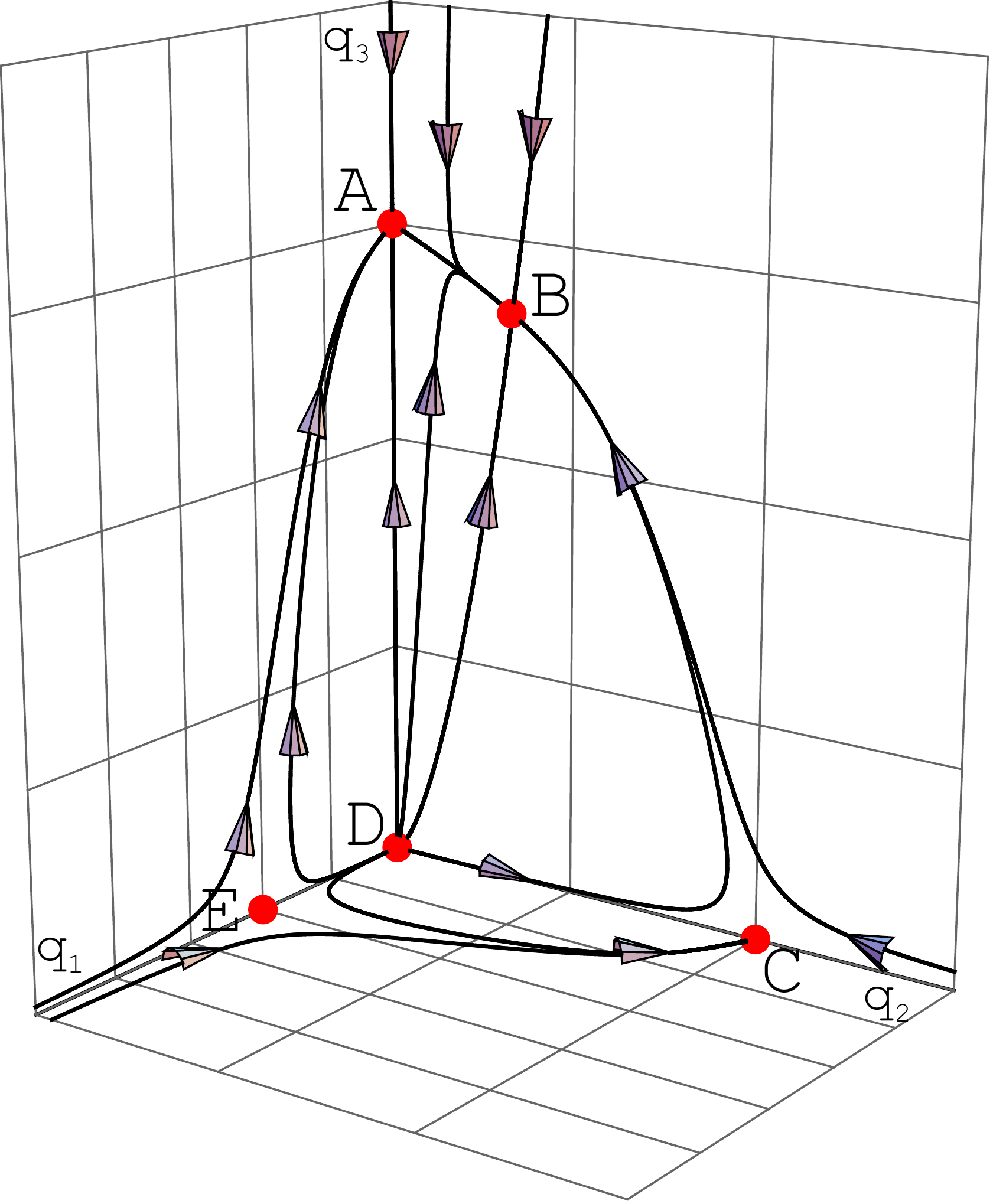}
    \caption{The co-existence of three two-dimensional submanifold of system ()
representing the planes of $\{(q_1,q_2),q_3=0\}$, $\{(q_1,q_3),q_2=0\}$ and
$\{(q_2,q_3),q_1=0\}$.}
    \label{fig:3}
\end{figure}

Let us consider the first system on invariant submanifold (\ref{eq:7_s1}) and
perform the dynamical analysis for it. This system has four critical points
\begin{align}
E_{1} &= (0,\ 0) \\
E_{2} &= \left( \frac{a_1}{2b_1},\ 0, \right) \\
E_{3} &= \left( 0,\ \frac{a_2}{2b_2}, \right) \\
E_{4} &= \left( \frac{2 a_1 b_2 - b a_2}{4 b_1 b_2 - b^2},\ 
\frac{2 a_2 b_1 - b a_1}{4 b_1 b_2 -b^2}, \right) \label{eq:34}
\end{align}
The phase portrait for this system is presented on the plane 
$\{ (q_2,q_3),q_1=0\}$ in Fig.~\ref{fig:3}.

In the positive quadrant there is only one critical point: the stable node,
which corresponds to the duopoly equilibrium. Only two firms exist on the
market and they reach the stable equilibrium with fixed production level. The
condition for the existence of this point is
\begin{equation}
2a_1 b_2 - b a_2 > 0 \quad \text{and} \quad 2a_2 b_1 - b a_1 > 0.
\end{equation}

For the chosen parameter values (\ref{eq:33}) we have $2a_2 b_1 - b a_1 = 0$.
In this case there is no critical point inside the positive quadrant as well as
on the respective submanifolds, which are the planes $\{ (q_1,q_3),q_2=0\}$ and 
$\{ (q_1,q_2),q_3=0\}$. Markets with initial conditions in those planes tend
towards monopolies on axes $q_2$ or $q_3$ as seen in fig.~\ref{fig:3}, the
exception being the $q_1$ axis which is an invariant submanifold itself.

\subsection{Lapunov function}

For the system (\ref{eq:7_s3}) we will look for the Lapunov function. We assume
that both conditions $2a_1 b_2 - b a_2 > 0$ and $2a_2 b_1 - b a_1 = 0$ are
fulfilled and $q_1^* > 0$ and $q_2^* >0$, so we consider the critical point
(\ref{eq:34}).

Let us take the Taylor expansion of the right-hand sides of (\ref{eq:7_s3}) in
the neighbourhood of the critical point $(q_1^*, q_2^*)$ given by (\ref{eq:34})
and write the system in the form \cite[p.~206-207]{Zhang:2005de}
\begin{equation}
    \dot{x} = A_{2\times 2}x + g(x)
\end{equation}
where $x$ is a vector of components $x_1 = q_1 - q_1^{*}$ and $x_2 = q_2 - q_2^{*}$. The matrix $A$ and the vector $g(x)$ has the form
\begin{equation}
    A = \left[ \begin{array}{cc}
    -2\alpha_1 b_1 q_1^*  & - \alpha_1 b q_1^* \\
    -\alpha_2 b q_2^*   & -2\alpha_2 b_2 q_2^*
    \end{array} \right] = 
    \left[ \begin{array}{cc}
    a_{11} & a_{12} \\ a_{21} & a_{22}
    \end{array} \right]
\end{equation}
where 
\begin{align}
p &= \tr A = -2(\alpha_1 b_1 q_1^* + \alpha_2 b_2 q_2^*) < 0 \\
q &= \det A = \alpha_1\alpha_2 q_1^* q_2^* (4b_1 b_2 - b^2) > 0
\end{align}
as $b_1 > b$ and $b_2 > b$, and
\begin{equation}
    g(x) = \left[ \begin{array}{c}
    -2\alpha_1 (2b_1 x_1^2 + b x_1 x_2) \\
    -2\alpha_2 (2b_2 x_2^2 + b x_1 x_2)
    \end{array} \right].
\end{equation}
We define the Lapunov function \cite[p.~206-207]{Zhang:2005de}
\begin{equation} \label{eq:38}
V(x) = x^T K x
\end{equation}
where $K$ is a $2\times 2$ constant symmetric matrix which is a solution to $A^T K + KA = - I_{2\times 2}$ in the form $K = m (A^T)^{-1} A^{-1} + nI_{2\times 2}$. It takes place if $m= -q/2p$ and $n = - 1/2p$. Hence, we find that
\begin{equation} \label{eq:39}
    K = - \frac{1}{2pq} \left[ \begin{array}{cc}
    a_{21}^2 + a_{22}^2 + q & - a_{11} a_{12} - a_{21} a_{22} \\
    - a_{11} a_{12} - a_{21} a_{22} & a_{11}^2 + a_{12}^2 + q
    \end{array} \right]
\end{equation}
And the Lapunov function
\begin{equation}
V = - \frac{(a_{22} x_1 - a_{12}x_2)^2 + (a_{21}x_1 - a_{11} x_2)^2 + q(x_1^2 + x_2^2)}{2pq} > 0 \quad \text{for} \quad x \ne 0
\end{equation}
is positive definite.

Given that the Lapunov function is of the form (\ref{eq:38}) and matrix $K$ (\ref{eq:39}) we get
\begin{equation}
    \dot{V} = - x_1^2 - x_2^2 + 2 g^T K x
\end{equation}
where
\begin{align}
2 g^T K x &= \frac{1}{|p|q} \{ -2\alpha_1 (2b_1 x_1^2 + b x_1 x_2) [(a_{21}^2 + a_{22}^2 + q)x_1 - (a_{11} a_{12} + a_{21} a_{22})x_2] \} \nonumber \\
\qquad & -2\alpha_2 (2b_2 x_2^2 + b x_1 x_2) \{ (- a_{11} a_{12} - a_{21} a_{22}) x_1 +  (a_{11}^2 + a_{12}^2 + q) x_2 \}.
\end{align}

For any $p$ and $q$, the first two terms $-x_1^2 - x_2^2$ predominate over the
third term $2g^T K x$ in the neighbourhood of the origin, and $\dot{V}$ is
negative definite. Therefore, system (\ref{eq:7_s3}) has a strict Lapunov
function.

\section{The first integral analysis}

For the purpose of the subsequent sections it will be convenient to adopt a
redefined set of parameters. As the system has the general form
\begin{equation}
   \frac{d q_i}{d t} = \alpha_i q_i\left(a-d_i - (b+2e_i)q_i - b Q\right),\qquad i=1,2,3, 
\end{equation}
where $Q:=q_1+q_2+q_3$, one can see that $b=0$ makes the equation decouple and
the system completely solvable. It thus makes sense to rescale the time by
$t\rightarrow t/b$ and introduce new parameters:
\begin{equation} \label{reparam}
\begin{aligned}
    f_i &:= \frac{\alpha_i}{b}(a-d_i),\\
    \epsilon_i &:= \frac{2\alpha_i}{b}(b+e_i). 
\end{aligned}
\end{equation}
We can then make use of the following matrix notation
\begin{equation}
\frac{dq_i}{dt} = q_i \left(f_i + \sum_{j=1}^3\beta_{ij} q_j\right), \label{eq:41}
\end{equation}
where
\begin{equation}
    \beta :=
    -\begin{bmatrix}
    \epsilon_1 & \alpha_1 & \alpha_1\\
    \alpha_2 & \epsilon_2 & \alpha_2\\
    \alpha_3 & \alpha_3 & \epsilon_3
    \end{bmatrix}.
\end{equation}

\subsection{A general result regarding analytic first integrals}

In a polynomial system it is natural to look for polynomial first integrals,
which are a special case of analytic ones, but even the latter
class is very restrictive. In particular, if one demands that the conserved
function $I$ be analytic in all variables, then each hyperbolic critical point
provides additional necessary conditions. For around it, the system can be
linearized to be
\begin{equation}
    \dot{q_i} = \lambda_i q_i + \mathcal{O}(q^2),\quad i=1,\ldots, 3,
\end{equation}
assuming the point is at $q=0$. The derivative of the lowest monomial in the
expansion of $I$ around that point is then
\begin{equation}
    \frac{\mathrm{d}}{\mathrm{d}t}(q_1^{n_1}q_2^{n_2}q_3^{n_3}) = 
    (\lambda_1 n_1+\lambda_2 n_2 +\lambda_3 n_3) (q_1^{n_1}q_2^{n_2}q_3^{n_3})
    +\mathcal{O}(q^{n_1+n_2+n_3+1}).
\end{equation}
It follows that for $\dot{I}$ to vanish, $\lambda_i$ must be linearly
dependend over $\mathbb{N}$. The general statement, due to Poincar\'e, is that
the same holds when $\lambda_i$ are the eigenvalues of the linearization matrix
around any critical point, even if there are Jordan blocks.

In our case this immediately leads to the stringent result:
\begin{thm}
If the system has a first integral analytic at the origin, the quantities
$f_1$, $f_2$ and $f_3$
must be linearly dependent over the non-negative integers.
\end{thm}
In particular this means that $f_i$ cannot all have the same sign bceause $n_1
f_1 + n_2 f_2 + n_3 f_3$ cannot vanish for any positive integers $n_i$. The cases
{\it 4.1} and {\it 4.2} are thus precluded from having an analytic integral.

The same can be done around all the other points, the most straightforward ones
being $E_2$, $E_3$ and $E_4$, for which we have the following sets of
eigenvalues
\begin{equation}
    \left\{-f_i,f_j-\frac{\alpha_j f_i}{\epsilon_i},f_k - \frac{\alpha_k
    f_i}{\epsilon_i}\right\},
\end{equation}
where $(i,j,k)$ are cyclic permutation of $(1,2,3)$. These triples must be
independent over $\mathbb{N}$ for there to exist a first integral analytic at
$E_2$, $E_3$ and $E_4$, respectively.

The drawback of this approach is that the analysis has to be done separately at
each point, and the sets of integer coefficients $n_i$ is different every time.
Also, the requirement of analyticity is rather restrictive and we shall see in
the next section that under fairly general assumptions there can still exist
rational first integrals.

\section{Darboux Polynomials}

The analysis of rational integrals for homogeneous systems relies on the so
called Darboux polynomials $F$, which are defined by the following property
\begin{equation}
    \dot{F} = \sum_{i=0}^3 V_i \partial_{x_i} F =: P F,
\end{equation}
where $P$ is called the cofactor, and is also necessarily a polynomial. The
polynomials $F$ are sometimes called partial or second integrals, since they
are not conserved in general but each of them is conserved on the set defined
by $F=0$ as seen above. If the initial conditions are such that the system
starts in such a set it will remain there for all later times. If the cofactor
is zero, then the polynomial is just a first integral.

The first fundamental fact that we will use here is that if the system has a
rational integral $R=R_1/R_2$ with $R_1$ and $R_2$ relatively prime, they must
both be Darboux polynomials with the same cofactor. This is immediately seen
from direct differentiation.

Secondly, any product of Darboux polynomials is itself a Darboux polynomial
\begin{equation}
    \frac{d}{d\tau}\left(\prod_{k=1}^n F_k^{\gamma_k}\right)
    =\left(\sum_{k=1}^n\gamma_k P_k\right)
    \prod_{k=1}^n F_i^{\gamma_k},\quad \gamma_k\in\mathbb{N}.
    \label{prod}
\end{equation}
Note that if the domain of motion is such that the above functions are well
defined, any real $\gamma_k$ are admissible in a construction of first
integrals, although it might not be rational or even algebraic then. In other
words, if there are enough Darboux polynomials, so that one can find numbers
$\gamma_k$ such that the cofactor in \eqref{prod} vanishes, a first integral
can be found.

Let us now apply the above to our system. It is immediately visible that each
of the dependent variables $q_i$ is itself a Darboux polynomial with cofactors
given by $f_i+\Sigma_j\beta_{ij}q_j=:f_i+L_i$. Their linear dependence is a
major factor in determining integrability.

Since these are three polynomials in three variables, they will generically be
linearly independent. In fact, dependence does not occur for economically
viable values of the parameters, but we include this case nevertheless for the
sake of completeness -- the result applies to any system of the prescribed
form.

\subsection{The case $\det(\beta)=0$}

If it so happens that 
\begin{equation}
    -\frac{\det(\beta)}{\alpha_1\alpha_2\alpha_3} = 
    2-\frac{\epsilon_1}{\alpha_1}-\frac{\epsilon_2}{\alpha_2}
    -\frac{\epsilon_3}{\alpha_3} 
    + \frac{\epsilon_1\epsilon_2\epsilon_3}{\alpha_1\alpha_2\alpha_3}
    =0, 
\end{equation}
a linear combination of $L_i$ will vanish and this will lead to at least a
time-dependent first integral. To see this, let us notice that when the
determinant of $\beta$ is zero, it has a null left eigenvector $\gamma_k$ which
is not the zero vector
\begin{equation}
    \sum_{k=1}^3\gamma_k \beta_{kj} = 0,\quad j=1,\ldots,3.
    \label{null}
\end{equation}
The coefficients $\gamma_k$ might not be integer, but in any case the following
function will be defined  for positive $q_j$:
\begin{equation}
    I_0 := \prod_{j=1}^3 q_j^{\gamma_j},
\end{equation}
whose cofactor, by \eqref{prod} will be
\begin{equation}
    P_0 = \sum_{j=1}^3 \gamma_j \left(f_j + \Sigma_{k=1}^3\beta_{jk}q_k\right) 
    = \sum_{j=1}^3 \gamma_j f_j.
\end{equation}
Because this is constant, the equation $\dot{I_0}= P_0 I_0$ can be immediately integrated to yield the time-dependent integral
\begin{equation}
    \mathrm{e}^{-P_0 t}I_0 = \mathrm{const.}
\end{equation}

It might additionally happen that $P_0=0$, so there is no time dependence, or
that there are two zero eigenvalues of $\beta$ such that two independent null
vectors $\gamma_k$, $\theta_k$ can be found. In the second case there will be
two first integrals and time can be eliminated from at least one of them,
because both depend on it exponentially.

To illustrate the above, let us look at the case when all $\alpha_i$ are equal.
This is not a crucial assumption in this case, since they must all be non-zero,
so their values do not change the condition that $\det(\beta)=0$. The other
parameters must satisfy some constraint for the condition to be true, and there
are many possibilities due to the large number of parameters, but equation
\eqref{null} can always be solved, For example, in the generic case of
$\gamma_k\neq0$, the following parameters give vanishing determinant:
\begin{equation}
    e_i=-b\frac{\gamma_1+\gamma_2+\gamma_3+\gamma_i}{2\gamma_i},
\end{equation}
The corresponding the first integral is
\begin{equation}
I = e^{-(f_1\gamma_1+f_2\gamma_2+f_3\gamma_3)t}
\,q_1^{\gamma_1}q_2^{\gamma_2}q_3^{\gamma_3}.
\end{equation}

After taking logarithm both sides and differentiating with respect to time we
obtain that the rates of growth are linearly dependent and satisfy the following condition
\begin{equation}
f_1 \gamma_1 + f_2 \gamma_2 + f_3 \gamma_3 = \gamma_1 \frac{\dot{q_1}}{q_1} +
\gamma_2 \frac{\dot{q_2}}{q_2} + \gamma_3 \frac{\dot{q_3}}{q_3}.
\end{equation}
The rates of growth of firms' production are strictly related as a given firm
need to adjust its production rate of growth with respect to production rates
of growth of two other firms. This relation is a dynamical constraint imposed
on the production rates of growth at any moment. It is analogous to the
reaction curves analysis \cite{Puu:2011ooe}.

In a special case of constant rate of growth there is a solution 
\begin{equation}
q_1(t) \propto e^{f_1 t}, \quad q_2(t) \propto e^{f_2 t}, 
\quad q_3(t) \propto e^{f_3 t}.
\end{equation}

\section{The case $\det(\beta)\neq0$.}

We can next proceed to the generic, and more difficult, case when
$\det(\beta)\neq0$, noting first that any cofactor of our system must be of the
form
\begin{equation}
    P_4 = p_0 + \sum_{i=1}^3 p_i q_i,
\end{equation}
because the right-hand sides of the system are quadratic.

We will first try to find all linear Darboux polynomials
\begin{equation}
    F_4=w_0 + \sum_{i=1}^3w_i q_i,
\end{equation}
which do not reduce to just $q_i$ or a constant. Calculating $\dot{F_4}-P_4
F_4$ and equating coefficients of different monomials to zero, gives  system of
polynomial equations for $w_i$ and $p_i$:
\begin{align}
&q^0:  &w_0 p_0 = 0,\\
&q^1:  &\left\{\begin{aligned}
w_0 p_1+w_1(p_0-f_1) &= 0,\\
w_0 p_2+w_3(p_0-f_2) &= 0,\\
w_0 p_3+w_3(p_0-f_3) &= 0,
\end{aligned}\right.\label{lin}\\
&q^2:  &\left\{\begin{aligned}
w_1(p_1+\epsilon_1) &= 0,\\
w_2(p_2+\epsilon_2) &= 0,\\
w_3(p_3+\epsilon_3) &= 0,\\
w_1(\alpha_1+p_2)+w_2(\alpha_2+p_1) &=0,\\
w_2(\alpha_2+p_3)+w_3(\alpha_3+p_2) &=0,\\
w_3(\alpha_3+p_1)+w_1(\alpha_1+p_3) &=0,
\end{aligned}\right.
\end{align}


In the generic situation all $w_i$ are non-zero, which then leads to a homogeneous linear system:
\begin{equation}
\begin{bmatrix}
    \alpha_1-\epsilon_2 & \alpha_2 - \epsilon_1 & 0\\
    \alpha_1-\epsilon_3 & 0 & \alpha_3 - \epsilon_1\\
    0 & \alpha_2-\epsilon_3 & \alpha_3 - \epsilon_2
    \end{bmatrix}
    \begin{bmatrix}w_1\\w_2\\w_3\end{bmatrix}=0,
\end{equation}
whose non-zero solution requires the determinant to vanish
\begin{equation}
    D_1:=(\alpha_1-\epsilon_3)(\alpha_2-\epsilon_1)(\alpha_3-\epsilon_2)
    +(\alpha_1-\epsilon_2)(\alpha_2-\epsilon_2)(\alpha_3-\epsilon_1) = 0.
\end{equation}

If a solution of the system exists, the set of equations \eqref{lin} gives three constraints
\begin{equation}
    f_i = p_0 -\frac{w_0}{w_i}\epsilon_i,
\end{equation}
which have to take into account that $w_0 p_0=0$. So either $w_0=0$ and
$p_0=f_i$, which must all be equal; or $w_0=1$ ($F$ is always defined up to a
constant factor), $p_0=0$ and $f_i w_i = \epsilon_i$.

The remaining exceptional cases occur when one of $w_i$ is zero, and because
the system has complete symmetry with respect to the indices, it is enough to
consider $w_1=0$. Direct computation then shows that the only non-trivial
solution is
\begin{equation}
    F_4 = (\alpha_2-\epsilon_2)q_2 + (\epsilon_3-\alpha_2)q_3,\quad
    P_4 = f_2 - \alpha_2 q_1 -\epsilon_2 q_2 - \epsilon_3 q_3,
\end{equation}
with the additional constraints of
\begin{equation}
    f_2 = f_3,\quad \alpha_2 = \alpha_3.
\end{equation}
All the other solutions (and constraints) can be obtained by cyclic
permutations of the indices: $1\rightarrow 2\rightarrow 3\rightarrow 1$.

One should next proceed with all higher-degree Darboux polynomials, or, ideally
solve the problem for general degree $n$. Direct computation shows that there
are no new second or third order polynomials, other than products of all the linear
ones. The authors have been unable to proceed with the proof for the general
case, i.e., when all the parameters are independent. At the same time, imposing
some restrictions immediately produces linear Darboux polynomials and first
integrals, so it seems reasonable to formulate the following

\begin{conj} All Darboux polynomials of the system in question
    for general values of parameters are generated by the linear polynomials.
\end{conj}

So far we have not made use of the $\det(\beta)\neq0$ condition, and if Darboux
polynomials can be found via the procedure above, each of them defines an
invariant set $F=0$ by itself. But now that the matrix $\beta$ is not singular,
the linear forms $L_i$ form a basis, so the linear part of the cofactor $P_4$
can be decomposed in full analogy with the previous section:
\begin{equation}
    L_4 = \sum_i p_i q_i = 
    \sum_i \gamma_i L_i,\quad \gamma_i = \sum_j p_j [\beta^{-1}]_{ji}.
\end{equation}
Then the function
\begin{equation}
    I_0:=F_4 \prod_i x_i^{-\gamma_i},
\end{equation}
has a constant cofactor
\begin{equation}
    \dot{I_0} = \left(p_0 - \sum_i f_i \gamma_i \right)I_0=:P_0 I_0,
\end{equation}
and a time-dependent first integral is, as before, $I = \mathrm{e}^{-P_0 t}I_0$.

Coming back to the simpler cases of the system (\ref{eq:7}), we can consider all
firms identical i.e. their cost function are the same and/or the speed of
adjustment to market $\alpha_i$ are the same.

\subsection{Identical firms with linear or quadratic cost} 

The first two subcases, 2.2 and 2.4, of identical firms can in fact be treated together. That
is, regardless of whether the cost function is linear or quadratic we can find first
integrals.

Taking all $\alpha_i$ to be the same, and the cost function
\begin{equation}
C(q_i) = c + d q_i + e q_{i}^{2}, \quad c>0, d>0, e>0,
\end{equation}
the parameters (\ref{eq:20}) have the form
\begin{equation}
\begin{split}
f &= \frac{\bar{a}}{b} \\
\epsilon &= \frac{2}{b}(b+e)
\end{split}
\end{equation}
The matrix $\beta$ in this case is
\begin{equation}
    \det \beta = - \begin{vmatrix}
    \epsilon & 1 & 1 \\ 1 & \epsilon & 1 \\ 1 & 1 & \epsilon
    \end{vmatrix} = - 8 \left( \frac{b+e}{b} \right)^3  + 6 \frac{b+e}{b} - 2 \ne 0
\end{equation}
if $b>0$, $e>0$ so that $\frac{b+e}{e}>1$.

This is a degenerate case, because both $D_1$ vanishes {\it and} all $f_i$ are
equal. There are thus three additional linear Darboux polynomials, together
with their cofactors they are:
\begin{equation}
\begin{aligned}
    F_4 &= q_1 - q_2,& P_4 &= f - 2 q_1 - 2q_2 - q_3,\\
    F_5 &= q_2 - q_3,& P_5 &= f - q_1 -2 q_2 -2 q_3,\\
    F_6 &= q_3 - q_1,& P_6 &= f - 2 q_3 - q_2 - 2 q_1.\\
\end{aligned}
\end{equation}
According to the general treatment of section~8, we thus have three time dependent integrals
\begin{equation} \label{eq:83}
    \begin{aligned}
        I_4 = \mathrm{e}^{f t/4}q_3(q_1-q_2)(q_1q_2q_3)^{-3/4}, \\
        I_5 = \mathrm{e}^{f t/4}q_1(q_2-q_3)(q_1q_2q_3)^{-3/4}, \\
        I_6 = \mathrm{e}^{f t/4}q_2(q_3-q_1)(q_1q_2q_3)^{-3/4}, \\
    \end{aligned}
\end{equation}
and time can be eliminated from two of them, to yield a first integral
\begin{equation}
    I_1 = \frac{q_1(q_2-q_3)}{q_3(q_1-q_2)},
\end{equation}
and we note that cyclic permutations of indices produce also first integrals, but they are all functionally dependent.

\subsection{Firms with different quadratic terms of the cost functions}

This is a subcase of the situation of 2.4, and could be called ``almost
identical'' firms, as the parameters coincide in the linear parts ($d_i$ and
$\alpha_i$ are equal), but the quadratic terms of the cost functions are
different for each company. The previous two subcases are just restrictions of
this one, and we state here the general results regarding invariant
submanifolds and stability.

The parameters $f$ and $\epsilon_i$ (\ref{eq:20}) now have the form
\begin{equation}
\begin{split}
f_i &= \frac{a - d}{b}, \\
\epsilon_i &= \frac{2}{b}(b+e_i).
\end{split}
\end{equation}
The matrix $\beta$ in this case is
\begin{equation}
    \det \beta = - \left| \begin{array}{ccc}
    \epsilon_1 & 1 & 1 \\ 1 & \epsilon_2 & 1 \\ 1 & 1 & \epsilon_3
    \end{array} \right| = - \epsilon_1 \epsilon_2 \epsilon_3  + (\epsilon_1 + \epsilon_2 + \epsilon_3) - 2 .
\end{equation}
Because $b>0$ and $e_i > 0$ we have that $\epsilon_i > 2$ and hence $\det \beta < 0$.

There are again three additional Darboux polynomials
\begin{equation}
\begin{aligned}
    F_4 &= (\epsilon_1-1)q_1 - (\epsilon_2-1)q_2,& P_4 &= f - \epsilon_1 q_1 - \epsilon_2 q_2 - q_3,\\
    F_5 &= (\epsilon_2-1)q_2 - (\epsilon_3-1)q_3,& P_5 &= f - q_1 -\epsilon_2 q_2 -\epsilon_3 q_3,\\
    F_6 &= (\epsilon_3-1)q_3 - (\epsilon_1-1)q_1,& P_6 &= f - \epsilon_3 q_3 - q_2 - \epsilon_1 q_1.\\
\end{aligned}
\end{equation}
All the time dependent first integrals can then be obtained by cyclic index permutations from the following one
\begin{equation}
    I_4 = \mathrm{e}^{-P_0 t} q_3 \left((\epsilon_1-1)q_1-(\epsilon_2-1)q_2\right)
    q_1^{-\gamma_1}q_2^{-\gamma_2}q_3^{-\gamma_3},
    \label{int_4}
\end{equation}
where
\begin{equation}
    \gamma_i = \frac{\epsilon_i-1}{\det{\beta}}
    \left(1-\frac{\epsilon_1\epsilon_2\epsilon_3}{\epsilon_i}\right),
\end{equation}
and
\begin{equation}
    P_0 = \frac{(\epsilon_1-1)(\epsilon_2-1)(\epsilon_3-1)}{\det{\beta}}f.
\end{equation}
The elimination of time yields an ordinary first integral of
\begin{equation}
    I_1 = \frac{q_3((\epsilon_1-1)q_1-(\epsilon_2-1)q_2)}{q_1((\epsilon_2-1)q_2-(\epsilon_3-1)q_3)}.
\end{equation}

As mentioned before, each Darboux polynomial defines an additional invariant
submanifold $F_i=0$, which in this case are planes through the origin. In fact
all three intersect along a line because $F_6 = F_4+ F_5$ and their normal
vectors are not independent. Its equation is
\begin{equation}
    q =
    \left(\frac{q_4}{\epsilon_1-1},\frac{q_4}{\epsilon_2-1},
    \frac{q_4}{\epsilon_3-1}\right), \quad q_4\in(0,\infty).
\end{equation}
The Nash equilibrium $E_8$ lies exactly on this
line and is an attracting node.

We can say even more thanks to the explicit formula for $I_4$ and the
remaining two integrals. In the positive quadrant $q_1 q_2 q_3$ is nonzero,
while $P_0<0$ by the restrictions on the parameters. Thus solving the conservation
law $I_4=\text{const.}$ we immediately obtain from \eqref{int_4} that $F_4\rightarrow 0$ as
$t\rightarrow\infty$, and similarly for the other two.

This means that trajectories are attracted by all the invariant submanifolds
(planes)
and hence also by the line. On the line itself it is straightforward to check
that
\begin{equation}
    \dot{q_4} = q_4\left(f - \left(\frac{1}{\epsilon_1-1}+\frac{1}{\epsilon_2-1}+
    \frac{\epsilon_3}{\epsilon_3-1}\right)q_4\right)= f q_4(1-q_4/q_4^*),
\end{equation}
where $q_4^*$ is the position of $E_8$ in this coordinate. The Nash equilibrium
is thus seen to be the attractor of two trajectories
on the line, which in turn are attracting trajectories for the whole positive
quadrant.

\begin{figure}
    \centering
    \includegraphics[width=0.98\textwidth]{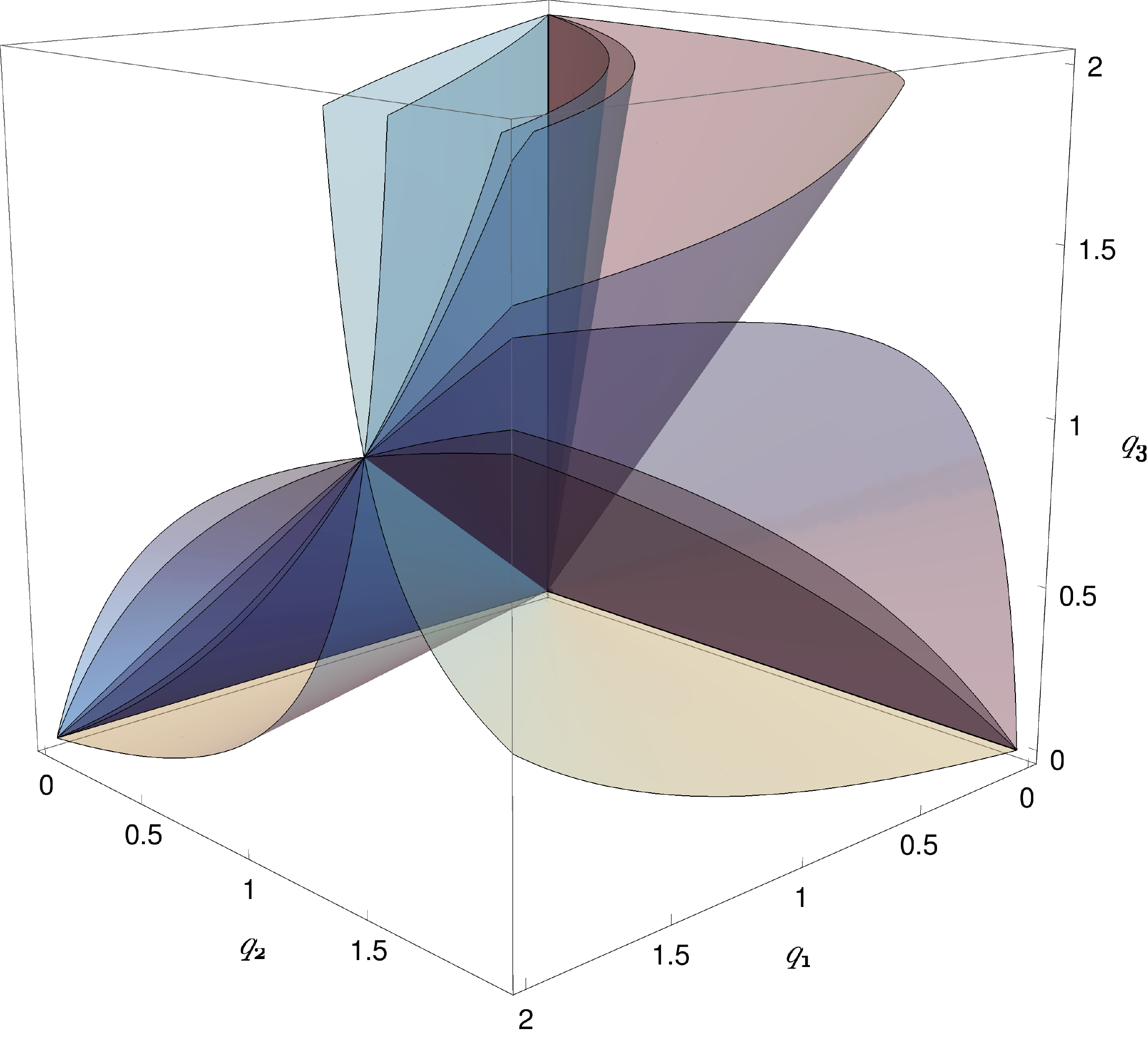}
    \caption{The level sets of the first integral $I_1$, for $\epsilon_1=2.1$,
    $\epsilon_2=2.5$, $\epsilon_3=2.9$.}
    \label{rozeta}
\end{figure}

The level sets of $I_1$, as seen in Figure \ref{rozeta} are surfaces also
intersecting at the
central line. Taking the first integral as one coordinate, the system becomes
just two dimensional on any such leaf, by solving $I_1=C$, eliminating one
of the original variables, e.g.
\begin{equation}
    q_3 = \frac{C(\epsilon_2-1)q_1 q_2}
    {(1-\epsilon_2)q_2+(\epsilon_1-1+C(\epsilon_3-1))q_1}=:Q_3(q_1,q_2),
\end{equation}
and substituting into the original system:
\begin{equation}
\begin{aligned}
    \dot{q_1} &= h^1\left(q_1,q_2,Q_3(q_1,q_2)\right)\\
    \dot{q_2} &= h^2\left(q_1,q_2,Q_3(q_1,q_2)\right).
\end{aligned}
\end{equation}

\subsection{Extension to $n$-dimensions}

The special shape of the Darboux polynomials and first integrals suggests a
straightforward generalisation to $n$ dimensions, and indeed it is easy to
check, that for $n$ almost identical (in the above sense) firms, one has a
family of Darboux $n(n-1)/2$ polynomials
\begin{equation}
    F_{ij} = (\epsilon_i-1)q_i - (\epsilon_j-1)q_j,\quad
    P_{ij} = f - \epsilon_i q_i -\epsilon_j q_j - 
    \sum_{k\neq i,j}q_k.
\end{equation}
Like before, there exist time-dependent first integrals
\begin{equation}
    I_{ij} = \mathrm{e}^{-P_0 t}F_{ij} \prod_k q_k^{-\gamma_{ij}^k},
\end{equation}
where $\gamma_{ij}^k$ is the $k$-th component of the vector
\begin{equation}
        \gamma_{ij} = -[1,\ldots,1,\epsilon_i,1,\ldots,
        1,\epsilon_j,1,\ldots,1]\beta^{-1},
\end{equation}
and it so happens that for all pairs $(i,j)$, the sum of the components is the
same giving
\begin{equation}
    P_0 = f-f\sum_k\gamma_{ij}^k = \frac{f}{\det\beta}\prod_{k}(\epsilon_k-1).
\end{equation}
The ratios of these integrals are thus time-independent, but again not all are
functionally independent. E.g., for $I_{123} := I_{12}/I_{23}$ and $I_{312} =
I_{31}/I_{12}$ we have
\begin{equation}
    I_{312} = \frac{1-\epsilon_1+I_{123}(1-\epsilon_3)}{I_{123}(\epsilon_2-1}.
\end{equation}

Finally, because $\det\beta<0$ all $F_{ij}$ tend to zero exponentially with
time, so all trajectories approach the line through the Nash equilibrium, as
before.

\section{Conclusions}

We used the qualitative methods of dynamical systems to study the phase
structure of the oligopoly model. Exploring the example of three firms
oligopoly we analyse the complexity of the model dynamics. The phase space is
organized with one-dimensional and two-dimensional invariant submanifolds (on
which the system reduces to monopoly and duopoly) and a unique stable node
(global attractor, Cournot and Nash equilibrium) in the positive quadrant of
the phase space.

Its inset is the positive quadrant of the phase space $\{ (q_1, q_2, q_3)\colon
q_1 > 0, q_2 > 0, q_3 > 0 \}$. The boundaries of this quadrant are three
two-dimensional submanifolds where the dynamics of duopolies is restricted to.
Trajectories from the bulk space depart asymptotically from unstable invariant
2-dimensional submanifolds. In a generic case they reach the Nash equilibrium
point. The dynamics of monopolies is restricted to the axes of coordinates
systems. On this line there exits stable critical points (A, C, F in Fig.~1)
they are a equilibrium points of monopolies.

From our dynamical analysis of the system one can derive general conclusion that
the system under consideration is structurally stable, because its phase
portrait in the generic case contains saddle and nodes. It seems to be important in
the economic context because its structural stability means that dynamics
cannot be destroyed by small perturbations \cite{Perko:2001de}.

The problem of integrability of the model was also addressed using Darboux
polynomials, since this is the natural setting for polynomial systems. Because
of the large number of parameters, the full analysis was not possible, but some
general cases of interest were shown to posses additional linear Darboux
polynomials and also time-dependent conserved quantities. This allowed to give
qualitative analysis of the asymptotic behaviour and dimensional reduction in
the general case of firms different at the quadratic level.

The following points of interest were also addressed:
\begin{itemize}
\item We formulate a criterion of asymptotic stability of a Cournot equilibrium
which indicate that phase space interior of positive octant in
$\mathbb{R}^3_{+}$. The phase space structure is organised as follows: the
global attractor and three repelling invariant submanifolds. They form the
boundary of this octant. From the economic point of view they represent the
dynamics of duopolies.
\item The Cournot equilibrium is a global attractor in the phase space for
generic class of initial conditions and model parameters.
\item We found the new algebraic interpretation of reaction curves for some
forms of oligopoly system admitting time dependent first integral. This
constraint assumes the form of linear combination of production rates of
growth.
\item We demonstrate how the dimension of the system can be lowered by one due
to existence of first integral in the case of firms with different quadratic
term in cost function.
\item When a first integral exists it extends our knowledge of invariant
submanifolds beyond the linear case (planes) as depicted in Fig.~4.
Additionally it points to the lack of chaotic behaviour in the system.
\item The relations between growth rates and the variables themselves obtained
thanks to the first integrals provide immediate observable constraints that can
be tested against data.
\end{itemize}

\section*{Acknowledgments}
This work has been supported by the grant No. DEC-2013/09/B/ST1/04130 of the National Science Centre of Poland. The authors thanks Franciszek Humieja for comments and remarks.

\section*{Bibliography}


\end{document}